# The pagoda instability (PI) on soluble fibers

Jinhong Yang, Quanzi Yuan[*]

State Key Laboratory of Nonlinear Mechanics, Institute of Mechanics, Chinese Academy of Sciences, Beijing 100190, People's Republic of China

School of Engineering Science, University of Chinese Academy of Sciences, Beijing 100049, People's Republic of China

This paper presents a new kind of instability when inserting a soluble fiber into liquid. After wetting and dissolving the fiber by the liquid, the moving contact line (MCL) spontaneously loses stability. Because the sculpted shape from fiber looks like a Chinese pagoda, we name this instability as pagoda instability (PI). Coupling of dissolution and wetting leads to other special phenomena, i.e. dissolving-induced jet flow, and optimizes the fiber shape, etc. We propose a criterion of PI and show the competition between interface energy and chemical potential deduce the MCL motion and PI. A phase diagram is used to summary the final shapes of fibers. By conducting atomic force microscope (AFM) measurement, we find the fiber with optimized-shape has the characteristics of low adhesion force. Using the optimized-fiber can decrease the 70% influence of capillary force for AFM measurement in humid environment.

The study on the behaviors and mechanisms for wetting on insoluble fibers is critical importance in the field of directional droplet transport[1,2], coating[3], surface modification of fibers[4], preparation of micro-/nano-devices[5], etc. Wetting should be treated as inherently irreversible processes, because is the process of energy dissipation[6]. Therefore, the instability of wetting requires the input of the external energy. For example, a lyophilic fiber is inserted into the liquid, the surface tension drives the liquid to rise along the fiber until the contact angle is equilibrium angle. The final height of moving contact line (MCL) $H \sim R_0 \ln(2\kappa^{-1}/R_0)$, where $R_0$ is the fiber radius, $\kappa^{-1} = \sqrt{\gamma/\rho g}$ is the capillary length, $\rho$ is the density, and $g$ is the gravitational acceleration[7]. The instability of MCL on the fiber requires exerting force to make the fiber left from the liquid. Once the fiber is pulled out from the liquid, Plateau-Rayleigh instability (PRI) happens. The liquid film on the fiber surface breaks into periodically-distributed droplets. The distance between neighboring droplets is proportional to the initial radius $R_i$ ($R_i = h_l + R_0$, where $h_l$ is the initial thickness of the liquid film.). PRI is dominated by surface tension, which minimizes the surface area of the liquid and the surface energy of the wetting system[8].

However, inserting a soluble fiber into liquid makes new phenomena immerge, which cannot be explained by the previous theory of wetting on the insoluble fiber[9,10]. Both surface tension and chemical potential gradient drive the liquid to rise on the surface of the fiber[11]. Once the MCL reaches specific heights, a new instability occurs spontaneously (Fig. 1). The liquid falls quickly and does not break into a series of droplets. Because the sculpted shape from fiber looks like a Chinese pagoda[12], we designate this instability as pagoda instability (PI). PI optimizes the shape from the soluble fibers to be a special shape, which is similar to the shape of the eaves of pagoda and can be found in some substances with low surface energy, such as pollens[12], spiky particles[13].

In this paper, we observe PI by inserting soluble fibers into the liquid. We design a series of experiments to quantitatively investigate the influence of the physical/chemical properties and the initial shape from fibers on the interfacial behaviors. The results of Particle Image Velocity (PIV) and time-lapse photography explain the dissolving-induced jet flow and the coupling of the fiber dissolution and capillary rise. We clarify the reason why the fibers dissolve into different shapes and reveal the dynamic rules and mechanisms for the evolution of interface and energy, as well as obtain the criterion of PI. Based on our results, we can predict the occurrence of PI and the final shape from the dissolved fiber. At the end of this paper, we use an atomic force microscope (AFM) to test the surface adhesion force of the fiber which is optimized by PI.

Inserting soluble fibers with different radiuses (0.5mm~5mm) into the liquid induces PI and forms various structures. The water rises along the fiber wall to a certain height, and then the water-oil interface falls to a new position and rises again. This loop repeats until the liquid separates from the fiber (see Fig. 1a). Every loop of the rise and fall is similar, so we propose that the forming process of pagoda structure is self-similar. Therefore, after clarifying the forming mechanism for the first floor, we can deduce the whole process of fiber dissolving. The profile of the first floor is shown in Fig. 2(a). Without loss of generality, we build a model which the fiber is surrounded by two liquids. One is solvent (the symbol is $L$), the other is covering that is immiscible with solute and solvent (the symbol is $O$). If the outside factors can be ignored, the covering does not need in the experiments. Then, the symbol $O$ represents air. The dissolution on the surface of the fiber results in instability, and the forming process of the pagoda structure can be divided into three stages: liquid rise, instability, and liquid fall.

Unbalanced forces cause the liquid to rise. We analyze the driving forces and resistances which arise from free energy and dissipation respectively. Free energy consists of interface energy and chemical potential results from dissolution and rise. With the change in interface shape induced by dissolution and rise, the direction of surface tensions and the interfaces vary, and then interface energies change. It is noted that the influence of concentration on surface tensions is so light that can be ignored based on the results of experiments[14]. When the solute enters the liquid from solid, solvent molecules surround it[15]. Therefore, the chemical potential is related to the construction of solute molecules except the concentration. We use solvation energy density $\Gamma$ to characterize the solute influence to solution[16]. The dissolution effect can be expressed as a force with surface tension dimension $2\Gamma \bar{c} \bar{\delta}_{SL}$, where $\bar{\delta}_{SL}$ and $\bar{c}$ are the characteristic thickness of the boundary layer near the solid-liquid ($SL$) interface and averaged concentration in the boundary layer, respectively. The dissipation is induced by contact line "friction" and liquid viscosity[17]. The contact friction can be express as $\pi \xi_{cl} R_n \dot{h}$, where $\xi_{cl}$ is friction parameter, $R_n$ is the n$^{th}$ floor initial radius, $\dot{h}$ is the liquid rise velocity. The viscosity dissipation mainly exists in the boundary layers of liquid L and liquid O. The viscosity dissipation is expressed as $3\pi\eta R_n \delta^{-1} \int_0^h v^2 dz$, where $\eta$ is the liquid viscosity, $\delta$ is the boundary layer thickness, $v$ is the flow velocity near liquid $L$-liquid $O$ ($LO$) and $v \sim \dot{h}$. The boundary layer thickness is proportional to $\sqrt{\eta R_0 / \rho v}$ where $\rho$ is the density of the liquid. We ignore the gravity effect on the $LO$ interface based on the low Bond number (Bo < 0.1). Because the rise velocity is slow, we hypothesize the process of rise is quasi-static. In other words, the driving forces always keep balance with resistances. Consequently, based on Onsager's variational principle and ignoring the change in capillary number[18], we obtain the scaling law of liquid rise

$$\frac{h}{R_n} = \left(\frac{t}{\tau}\right)^{1/3} \qquad (1)$$

Here, $\tau = \left(C_\gamma - 2\xi_{cl}\right)/\left(R_n^3 C_a^2\right)$ is the characteristic time, and $C_\gamma$, $C_a$ are constants. $C_\gamma = \left(\gamma_{SO} - \gamma_{SL}\cos\theta_{SL} + \gamma_{LO}\cos\theta_{LO} + 2\Gamma\bar{c}\delta_{SL}\right)/\dot{h}$, where $\gamma$, $\theta$ and the subscript $SO$ are surface tension, contact angle, and solid-liquid $O$ interface, respectively. $C_\gamma$ shows that the dissolution effect can be expressed as the generalized surface tension and influence the wetting. Liquid rises with the constant advancing angle[19] $\theta_A$ that is the sum of $\theta_{SL}$ and $\theta_{LO}$. With the increasing of contact angle $\theta_{SL}$, the contact angle $\theta_{LO}$ decreases, and the rise velocity of the liquid slows down.

The liquid rises with the recession of the $SL$ interface. As shown in Fig. 2(b), the dissolution takes place in boundary layers near the $SL$ interface and results in dissolving-induced jet flow. The thickness of the boundary layer depends on the viscosity $\eta$ and diffusivity $D$. The relative importance of the viscosity and the diffusivity can be characterized by Schmidt number $\text{Sc} = \eta/(\rho_l D)$, and we estimate $\text{Sc} \sim 10^3$ in most cases. Therefore, the big Schmidt number implies that solid dissolves into solvent by shear effect rather than diffusion. The dissolution velocity of radial direction depends on the shear stress in the boundary layer $\delta_{SL}$, i.e. $v_n \propto |\tau| \sim \eta v_g / \delta_{SL}$ [20], where $v_g$ is the flow velocity near the $SL$ interface. Here, the boundary layer thickness near $SL$ interface $\delta_{SL} \propto \bar{\delta}_{SL}\sqrt{(1-z/h)} = \sqrt{\eta R_n (1-z/h)/(\rho_c U_N)}$, where $\rho_c$ is the characteristic density of liquid in this boundary layer, $U_N$ is the characteristic flow velocity near the $SL$ interface. The buoyancy differences in fluid induce the near-body flows and develop the interface layer[21]. Therefore, the characteristic flow velocity near the *SL* interface $U_N = \sqrt{\beta g c_s R_n}$, where $g$, $\beta$, $c_s$ are the gravitational acceleration, the solutal expansion coefficient, the saturation concentration of the solute, respectively[22]. The velocity can be expressed as $v_g = U_N \sqrt{h-z}/R_n$. The function can describe the results of raw data well (Fig. 3b). As we said before, *SL* interface recedes with the rise of liquid. Thus, the dissolution time is dominated by liquid rise, then we can obtain dissolution time is $t_d \propto (h^3 - z^3)$, according to the Eq. (1). Obviously, the profile of the *SL* interface can be represented as $f_{SL} = v_n t_d$. We assume the dissolution of solid below the horizon ($z < 0$) is uniform, because the slope of the fiber wall is small (<5°). Therefore, we think the shape from fiber below the horizon keeps cylindrical, so dissolution time and dissolution velocity satisfy $t_d \propto h^3$ and $v_n|_{z=0}$, respectively. It is worth explaining that the dissolution in the zone below the horizon is different, because the flow separation occurs at the end of fiber. There is a quiescent region on the bottom of fiber because the flow separates from the end of the fiber. In the quiescent region, the flow velocity is very slow and about $10^{-8}$ m/s. The transport method of solute depends on the relative importance of flow velocity and diffusion velocity, whose ratio is Péclet number. As we mentioned before, the diffusivity $D$ is about $10^{-9}$ m/s, and the size of the quiescent region $\delta_q$, is about $10^{-4}$ m, approximately, by experiment measurement. Péclet number is $10^{-2}$ and so low that the convection effect can be ignored in the quiescent region. The change in the fiber length can be expressed as $\Delta l = D(c_s - \bar{c}) t_d / \delta_q$. Besides, the profile of *LO* interface satisfies the equation $f_{LO}/\sqrt{1 + (df_{LO}/dz)^2} = R_n$. The condition $df_{LO}/dz|_{z=h} = \tan\left[\theta_A - \arctan\left(df_{SL}/dz|_{z=h}\right)\right]$ leads to the function of *LO* interface. According the Fig. 4, the key of pagoda construction formation is liquid fall caused by instability. Therefore, we further study the process of MCL instability.

According to the results of experiments (see fig. 5), the instability differs from PRI. From an energy perspective, the liquid rise is a process of free energy desecration. For the rise without dissolution, the *SL* interface replaces the *SO* interface whose interface energy is higher. During the rise of liquid on the surface of the soluble fiber, liquid rise accompanies dissolution. The *SL* interface replaces the *SO* interface while solute molecules enter the solvent further to reduce the free energy. As a result, the height of liquid rise accompanied dissolution is higher than that without dissolution. According to the above analyses, we built a model to calculate the change in the free energy *F* which is the function of liquid height *h* and includes the change in interface energy and chemical potential of the solution. The increment of *SL* interface energy $\gamma_{SL}\Delta A_{SL}$ always accompanies the decrease of *SO* interface energy $\gamma_{SO}\Delta A_{SO}$. Similarly, the change in interface energy in the *LO* interface can be written as the difference between the initial state $\gamma_{LO}\pi\left(f_{LO}|_{z\to 0}\right)^2$ and the instant state, i.e. $\gamma_{LO}\Delta A_{LO}$. As we said before, the dissolution occurs in the boundary layer $\delta_{SL}$ near the *SL* interface, the concentration of solution beyond the boundary layer $\delta_{SL}$ approaches zero. The average volume concentration in the boundary layer $\delta_{SL}$ is $c = v_n/v_g$. Thus, the change in chemical potential in solution resulting from dissolution is $\int \Gamma c \mathrm{d}V_{SL}$, where $V_{SL}$ is the volume of the boundary layer. The change in free energy in different stages is calculated by the numerical method and shown in Fig. 5(a). Obviously, with the rise of liquid along the fiber, the free energy has the minimum value, i.e. potential well. Elaborating on this process, the free energy decreases because the area of the *SL* interfaces increases resulting from the rise and dissolution of liquid. The dissolution keeps proceeding because the solution is unsaturated. The perturbation makes liquid oscillates around the potential well ($\mathrm{d}F/\mathrm{d}h = 0$, $\mathrm{d}^2 F/\mathrm{d}h^2 > 0$). We further calculate the changes in free energy during the liquid fall whose results are shown in Fig. 5(a) by blue lines. Free energy decreases with the fall of liquid. Given that, the position of the potential well is not the stable point of the system; the MCL losses stable in the position of the potential well. Fig. 5 shows that the liquid does not fall to the origin and stop in a certain position that we call it recovery of stability, unless the liquid separates from fiber during sliding. It is noted that the liquid falls with the receding angle that distinguishes with the advancing angle. In the process of liquid fall, the fall velocity is much higher than dissolution velocity, so we disregard the dissolution during liquid fall. During liquid *L* dissolving the solid, the *LO* interface losses stability as soon as liquid *L* reaches a certain height $H_n$ which is the height of the *n*[th] floor. By calculating the null point of derivative of the function *F* that the second derivative of *F* greater than zero, we can obtain the criterion of the interface instability

$$\frac{H_n^3}{\bar{L}_{SL} R^2} \propto \frac{\gamma_{SO}}{\varphi_c}, \qquad (2)$$

where the symbol $\varphi_c = \bar{\delta}_{SL} \Gamma \bar{c}$ is the dissolution energy per area. $\bar{L}_{SL}$ is the characteristic length of the *SL* interface. Consequently, the criterion can be express as $H_n \propto R_n^{2/3}$. PI is also the result of the competition between chemical potential and interfacial energy.

we disregard the dissolution effect during liquid falling, because the falling velocity is much quicker than the dissolution velocity. Therefore, the height of stability recovery can be expressed as

$$J_n \propto R_{n-1} \ln \frac{2\kappa^{-1}}{R_{n-1}}, \qquad (3)$$

which is similar with the height of the meniscus on the insoluble surface. The curves in Fig. 5(a) also show that the PI is periodic, that is the reason why solvent erodes the fiber into pagoda structures. According to the Fig. 5(a), the variations of free energy in each instability are similar, and that further proves that the fiber erosion is a self-similarity process.

We mention that the fiber can be eroded into different shapes including a single floor, pagoda-like, cuspidal tip, and flat end. Here, we elaborate on the forming process of the tip. During the formation of the tip, the dissolution velocity on the fiber wall (radial direction) competes with that in the bottom of fiber (axial direction). As we mentioned before, the convection effect dominates the dissolution on the fiber wall, and the diffusion effect dominates the dissolution on the bottom of fiber. Thus, in general, the dissolution velocity of the axial direction is slower than the radial direction. Unless the insertion depth of the fiber inserted is small, the fiber becomes a cuspidal tip. When the liquid separates from the fiber, or the dissolution depth in the radial direction is greater than the radius of fiber during the first instability, the fiber will format a single floor. Because of the axial dissolution of fiber, the inserted depth $l$ is the key parameter that decides the shape from the tip. According to the parameters $l$ and $R_0$, we divide all sizes of fibers into four types in the phase diagram (see fig. 5(c)). We use AFM to measure the adhesion force of reshaped fiber in contact mode, and the results of the AFM experiment are shown in Fig. 6. The adhesion force on the surface of the fiber is much smaller than the common commercial probe, because the special shape from tip helps to prevent the rise of the liquid like the eaves of pagoda expel the water. The PI-optimized fiber can be used to make the tip of the optical sensor[23], micro-needle patches for rapid and painless injection of drugs[24], super-hydrophobic surfaces that easily recover Cassie state[25], and so on.

In summary, we observe and explain the dynamic process of the soluble fiber inserted into liquid. The liquid rises along the soluble fiber to a certain height which is proportional to the initial radius to the power of 2/3, and then the MCL losses stability and falls. The competition of the dissolution and interface energy induces the moving and instability of MCL. In some cases, the liquid repeats the process of rise and fall, until the liquid separates from fiber completely. Because every instability is self-similar, the fiber forms a pagoda structure. We elaborate on the shape from fiber by phase diagram and measure the adhesion force of PI-optimized fiber which can shield the capillary force.

1. Bintein, P.B., Bense, H., Clanet, C. & Quéré, D. Self-propelling droplets on fibres subject to a crosswind. *Nat. Phys.* **15**, 1027-1032, (2019).
2. Zheng, Y. M. *et al.* Directional water collection on wetted spider silk. *Nature* **463**, 640-643, (2010).
3. Charles-Orszag, A. *et al.* Adhesion to nanofibers drives cell membrane remodeling through one-dimensional wetting. *Nat. Commun.* **9**, 4450 (2018).
4. Kaufman, J. J. *et al.* Structured spheres generated by an in-fibre fluid instability. *Nature* **487**, 463-467 (2012).
5. Panciera, F. *et al.* Controlling nanowire growth through electric field-induced deformation of the catalyst droplet. *Nat. Commun.* **7**, 1-8 (2016).
6. Yang, X. F. Equilibrium contact angle and intrinsic wetting hysteresis. *Appl. Phys. Lett.* 67, 2249 (1995).
7. De Gennes, P. G. Brochard-Wyart, F. & Quéré, D. *Capillarity and wetting phenomena: drops, bubbles, pearls, waves*. (Springer Science & Business Media, 2013).
8. Haefner, S. *et al.* Influence of slip on the Plateau–Rayleigh instability on a fibre. *Nat. Commun.* **6**, 7409 (2015).
9. Rayleigh, L. On the instability of a cylinder of viscous liquid under capillary force. *Philos. Mag. Sci.* **5**, 145-154 (1892).
10. Rayleigh, L. On the instability of jets. *P. Lond. Math. Soc.* **10**, 4-13 (1878).
11. Wang, G. & Lannutti, J. J. Chemical Thermodynamics as a Predictive Tool in the Reactive Metal Brazing of Ceramics. *Met. Mat. Trans. A* **26A**, 1499-1505 (1995).
12. Ito, S. & Gorb, S. N. Fresh "Pollen Adhesive" Weakens Humidity-Dependent Pollen Adhesion. *ACS Appl. Mater. Interfaces* **11**, 24691-24698 (2019).
13. Chen, H. J. *et al.* Anomalous Dispersion of Magnetic Spiky Particles for Enhanced Oil Emulsions/Water Separation. *Nanoscale* **10**, 1978-1986 (2018)
14. Docoslis, A., Giese, R. & Van Oss, C. J. Influence of the water–air interface on the apparent surface tension of aqueous solutions of hydrophilic solutes. *Colloids Surf. B* **19**, 147-162 (2000).
15. Yuan, Q. Z., Yang, J. H., Sui, Y. & Zhao, Y. P. Dynamics of Dissolutive Wetting: A Molecular Dynamics Study. *Langmuir* **33**, 6464-6470 (2017).
16. Yang, J. H., Yuan, Q. Z. & Zhao, Y. P. Solute transport and interface evolution in dissolutive wetting. *Sci. China Phys. Mech.* **62**, 124611 (2019).
17. Man, X. K. & Doi, M. Vapor-induced motion of liquid droplets on an inert substrate. *Phys. Rev. Lett.* **119**, 044502 (2017).
18. De Gennes, P. G. Wetting: statics and dynamics. *Rev. Mod. Phys.* **57**, 827-863 (1985).
19. Singler, T. J., Su, S., Yin, L. & Murray, B. T. Modeling and experiments in dissolutive wetting: a review. *J. Mater. Sci.* **47**, 8261-8274 (2012).
20. Moore, M. N. J. *et al.* Self-similar evolution of a body eroding in a fluid flow. *Phys. Fluids.* **25** 116602 (2013).
21. Wykes, M. S. D., Huang, J. M., Hajjar, G. A. & Ristroph, L. Self-sculpting of a dissolvable body due to gravitational convection. *Phys. Rev. Fluids* **3**, 043801 (2018).
22. Chong, K. L., Li, Y. S. Ng, C. S., Verzicco, R. & Lohse, D. Convection-dominated dissolution for single and multiple immersed sessile droplets. *J. Fluid Mech.* **892**, A21 (2020).
23. Tuniz, A. & Schmidt, M. A. Interfacing optical fibers with plasmonic nanoconcentrators. *Nanophotonics* **7**, 1279-1298 (2018).

24      Yu, J. C. et al. Microneedle-array patches loaded with hypoxia-sensitive vesicles provide fast glucose-responsive insulin delivery. Proceedings of the National Academy of Sciences of the United States of America 112, 8260-8265 (2015).
25      Li, Y. S., Quere, D., Lv, C. J. & Zheng, Q. S. Monostable superrepellent materials. Proceedings of the National Academy of Sciences of the United States of America 114, 3387-3392 (2017).

## Acknowledgements

This work was supported by the National Natural Science Foundation of China (Grant Nos. 11722223 and 11672300), the Chinese Academy of Sciences (CAS) Key Research Program of Frontier Sciences (Grant No. QYZDJ-SSW-JSC019), and the Strategic Priority Research Program of the Chinese Academy of Sciences (Grant No. XDB22040401),.

## Author contributions

## Competing interests

The authors declare no competing interests.

## Additional information

## Methods

**Experiment.** We used fibers made of caramel. Glucose was melt in 250 ℃ and pour into mold made by PDMS. We take the caramel out from mold when the caramel cools. The fiber was fixed in the lifting platform to control the position of fiber. We used water as the solvent, and the size of the container used to hold the water is 205×145×200 mm. Covering a thin layer of silicone oil can prevent the influence of humidity to experiment.

We use the Particle Image Velocity (PIV) technology to observe and record the flow field. A sheet light with 0.5 mm light waist was produced by a semiconductor laser. The polystyrene beads with 6 μm radius were mixed with water. The dissolving-induced jet flow and quiescent region was found.

In AFM experiments, the optimized-tip made of copper fiber was prepared by electrochemical corrosion. We cut the copper fiber with the optimized-tip and glue it with the AFM probe without tip. The test of adhesion was conducted in contact mode. We record the curve of force versus the distance between the tip and the surface of water.

**Governing equation of liquid rise.** Based on Onsager's variational principle, the governing equation of liquid rise is

$$\frac{\partial \text{Ca}}{\partial t} + \frac{\gamma_z}{\eta} = \frac{2\xi_{cl}}{\eta}\dot{h} + \frac{\text{C}_a}{\eta} h\dot{h}^{3/2}. \quad (4)$$

Here, $\text{C}_a = 7.5\sqrt{(\eta_a\rho_a + \eta\rho)/R_n}$, and $\gamma_z = \gamma_{SO} - \gamma_{SL}\cos\theta_{SL} + \gamma_{LO}\cos\theta_{LO} + 2\Gamma\bar{c}\bar{\delta}_{SL}$ is the resultant of surface tension in direction $z$. When the change in capillary number can be ignored, $\gamma_z$ is proportional to the rise speed $\dot{h}$, i.e. $\gamma_z = C_\gamma \dot{h}$ where $C_\gamma$ is constant. We rewrite Eq. (4) as

$$h\dot{h}^{1/2} + (2\xi_{cl}C_\gamma - C_\gamma^2)/C_a = 0 \text{ and obtain}$$

$$h = C_t t^{1/3}. \quad (5)$$

Here $C_t = \sqrt[3]{3(C_\gamma - 2\xi_{cl})^2/C_a^2}$.

**Free energy equation.** The change in the free energy can be written as

$$F = \gamma_{SL}\Delta A_{SL} + \gamma_{SO}\Delta A_{SO} + \gamma_{LO}\Delta A_{LO} + \int \Gamma c \mathrm{d}V_{SL}. \quad (6)$$

Considering the profile function of interfaces, eq. (6) can be further expressed as

$$F = \gamma_{SL}\pi\left[2\int_0^h (R_n - f_{SL})\mathrm{d}h + 2(R_n - v_n t_d)(l - \Delta l) - 2R_n l + (R_n - v_n t_d)^2 - R_n^2\right]$$

$$+ \pi\gamma_{LO}\left[2\int_0^h f_{LO}\mathrm{d}h - (f_{LO}|_{z=0})^2\right] - 2\pi\gamma_{SO}R_n h \quad (7)$$

$$+ 2\pi\Gamma\left[\int_0^h c(R_n - v_n t_d)\delta_{SL}\mathrm{d}h + \int_0^{l-\Delta} c(R_n - v_n t_d)\bar{\delta}_{SL}\mathrm{d}h + \frac{c_s}{2}(R_n - v_n t_d)^2 \delta_q\right]$$

Here, the dissolution time $t_d$ depends on the rise of liquid, i.e. $t_d = C_t^{-3}(h^3 - z^3)$.

## Data availability

The data that support the plots within this paper and other findings of this study are available in the main text and Supplementary Information. Additional information is available from the authors on reasonable request.

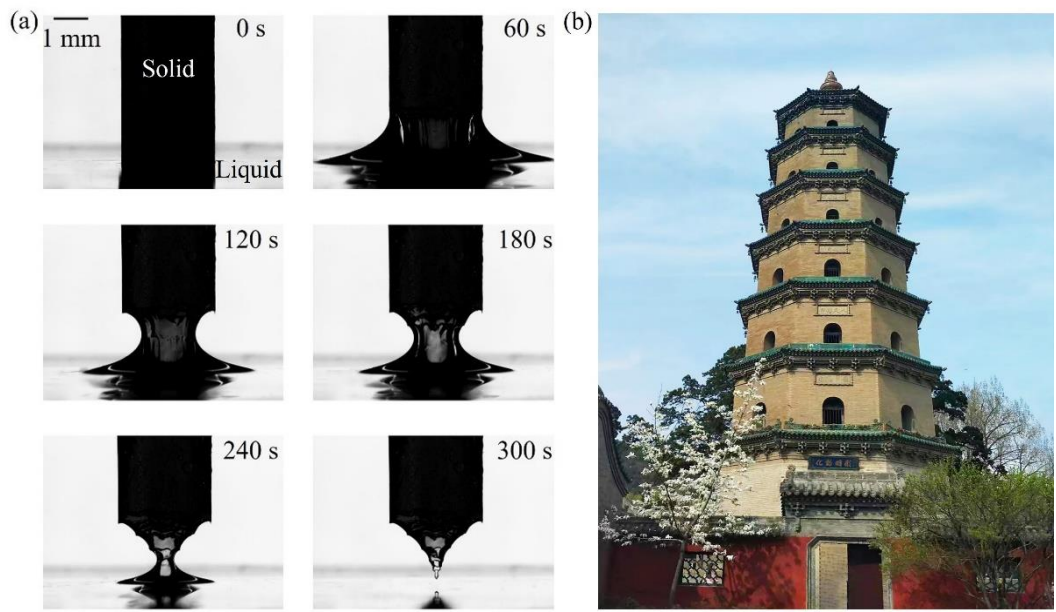

**Fig. 1 | The process of pagoda instability. a**, the Time-lapse photography of the process of the pagoda instability; **b**, Chinese pagoda.

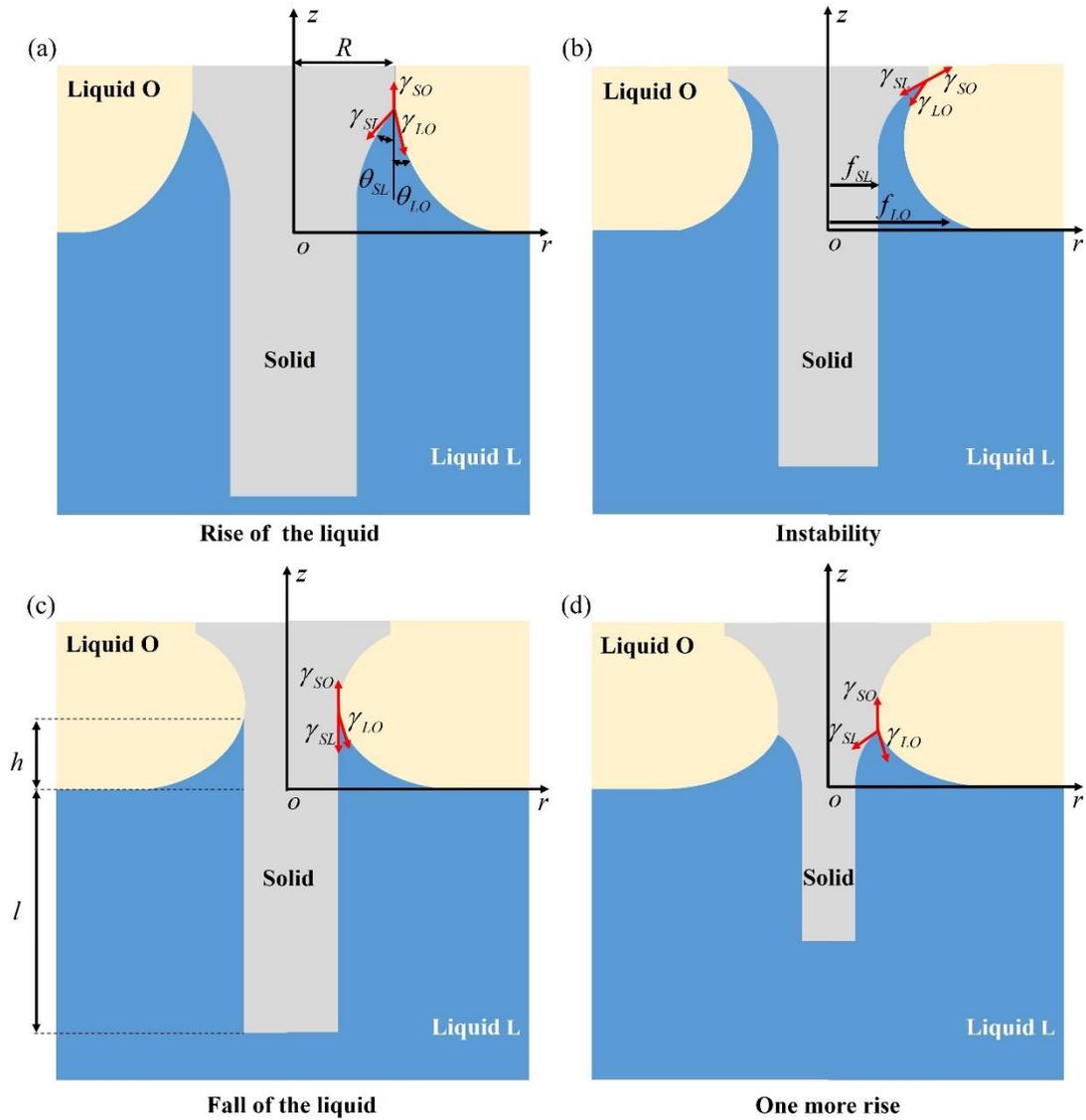

**Fig. 2 | Schematics of the model. a**, Liquid rises along the soluble fiber while liquid L dissolves solid. When the liquid rises to the certain position $H_n$, the $LO$ interface bends to fiber. **b**, contact line occurs instability. **c**, After instability, liquid L falls to new position. At last, the liquid rises again and enter next loop until the liquid separates with fiber. **d**, the liquid rises again.

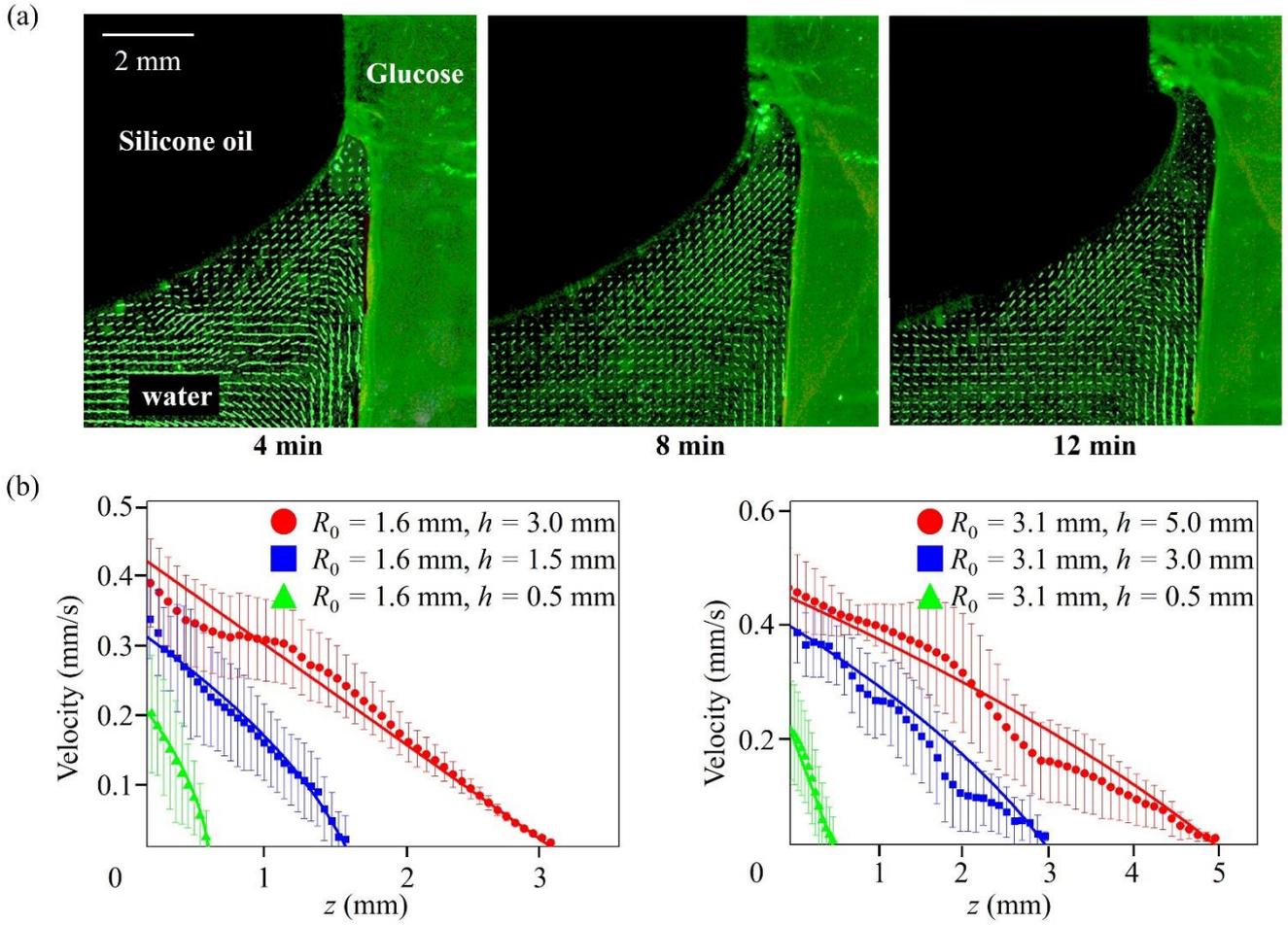

**Fig. 3 | Flow field. a,** The low concentration liquid flows up along Water-Silicone oil interface and dissolves the glucose. The high concentration liquid containing glucose flows down while the *SL* interface recedes. **b**, Plots shows that the velocity near the *SL* interface $v_g$ is proportion to $h$ and inversely proportional to $R$.

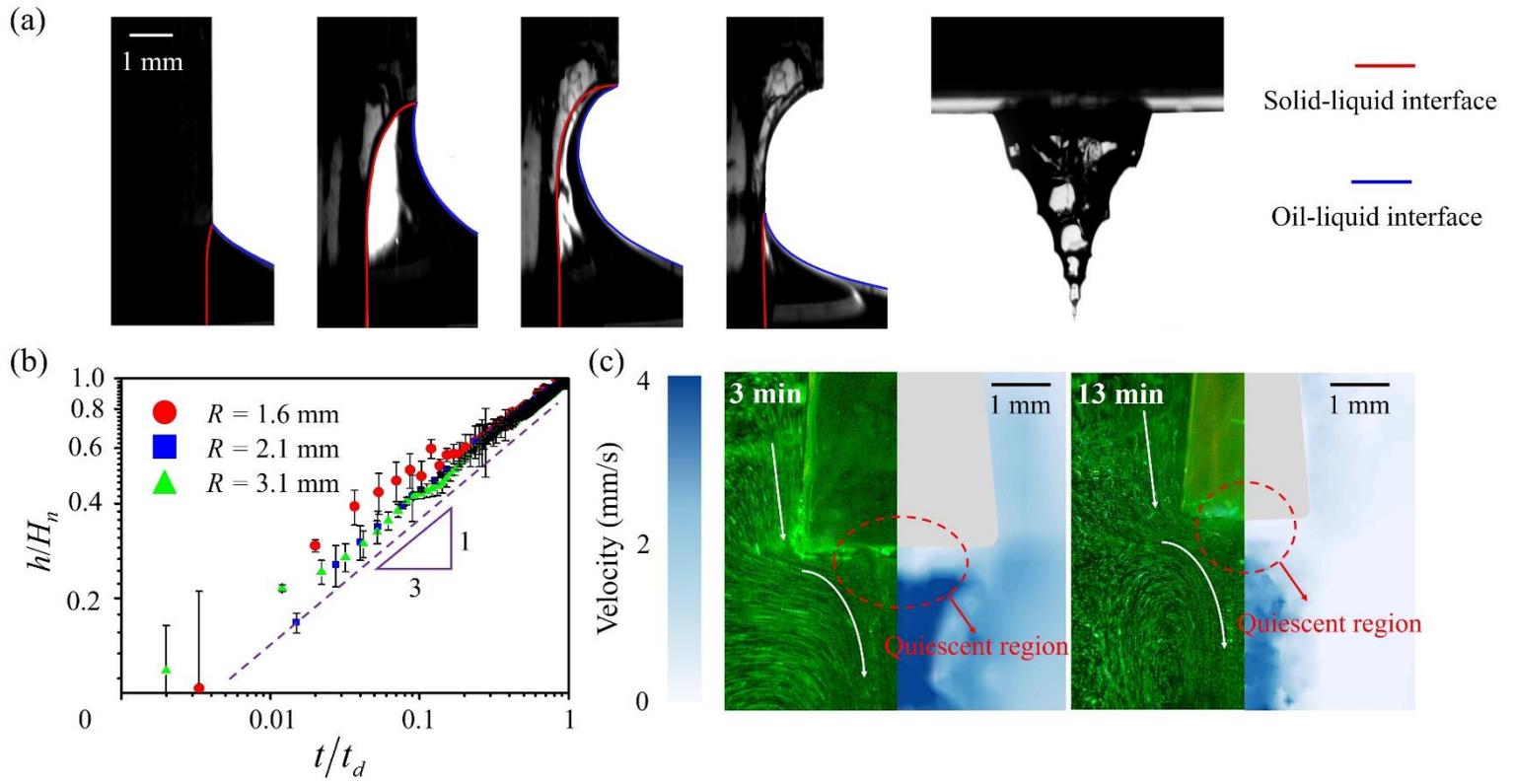

**Fig. 4 | Interface evolution of a dissolving fiber. a,** The time-lapse photos show the change at interface. Red lines and blue lines are the *SL* and *LO* interfaces which are obtained by theory. **b,** Plot shows the change in height $h$ with time. $t_d$ is the time taking to rise. **c,** PIV experiment indicate that the quiescent region exists in the bottom of fiber.

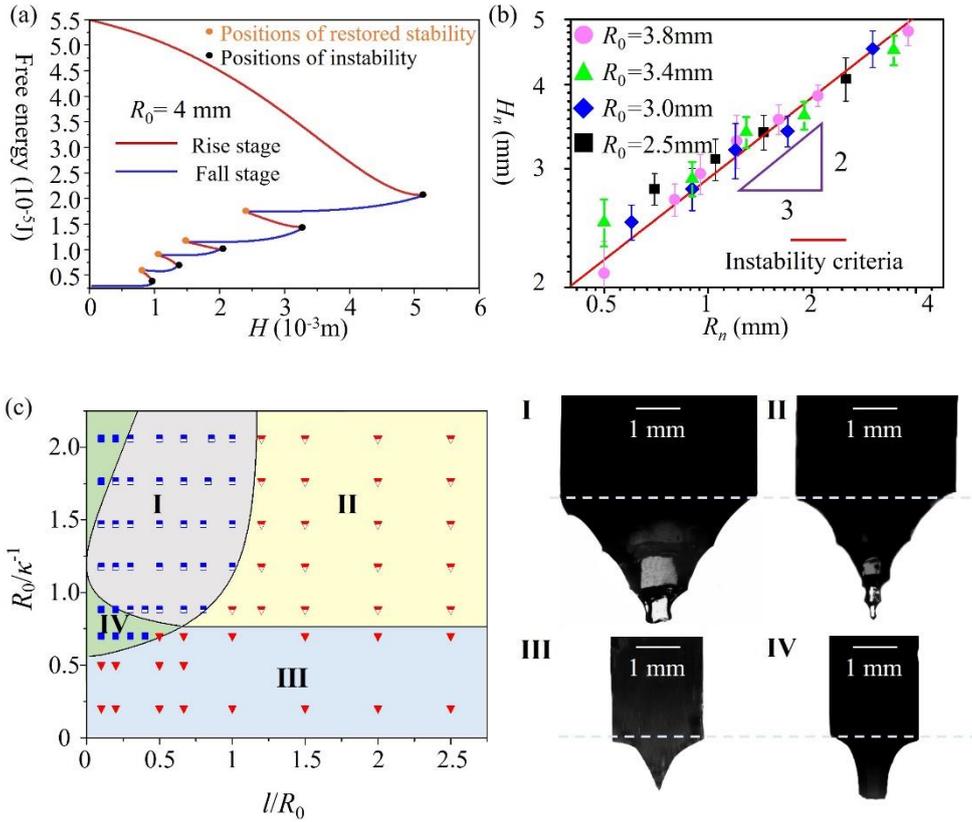

**Fig. 5 | Criterion of pagoda instability and shapes from dissolving fibers. a,** The diagram of free energy shows the change in energy in every loop (one rise and one fall) is similar. **b,** The instability criterion presents the scaling law of $H_n$ changing with $R_n$ is 0.5. **c,** Diagram shows the geometrical conditions obtaining different shapes fibers that include pagoda-likes with flat end, pagoda-likes with cuspidal tip, single floor with cuspidal tip, single floor with flat end (light gray,

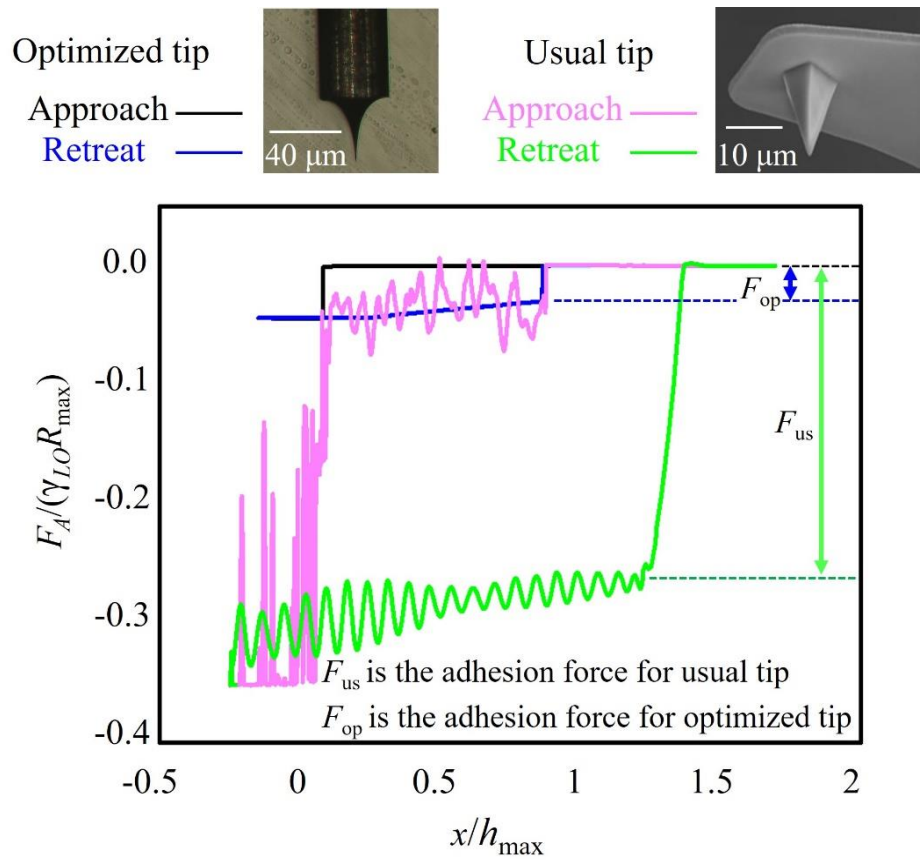

**Fig. 6 | Measuring adhesion force by AFM.** The black and pink lines are the loading curve of optimized tip and usual tip, respectively. The blue and green lines are the uploading curve of optimized tip and usual tip, respectively.